# Focusing limits of channeled high energy particles in bent crystals

Gennady Kovalev

*Wavecrest Corp., 7626 Golden Triangle Dr. Eden Prairie, MN 55344, USA*

**Abstract**

The focusing of high energy charged particles by a bent crystal is considered. The limits and constraints of focusing imposed by dynamics of particles are pointed out. The crystalline geometry provided the maximum magnification and minimum focusing size is proposed. The focusing parameters are compared with current experimental results.

## 1 Introduction

The possibility of beam steering of high energy particles by a crystal has been investigated since Tsyganov first proposed bent crystal channeling[1]. After the successful demonstration of the bent channeling[2], it became clear that bent crystals can be used for constructing a focusing element with extremely short focal length and focal spot close to size of one atomic channel. Several crystal devices for focusing were suggested [3–6]. Their common idea is based on differently bent plane channels to provide a 1-D beam convergence to a focal point. Because the crystal planes are naturally parallel, the focusing device must exert an external force making channels converged. Simple solution is a monocrystal partially cut into thin parallel slices with empty gaps between them[3]. To focus the beam the slices should be pressed together. Another solution is to apply a strong external pressure providing channel's convergence of the bulk crystal[5]. The further development in this direction led to mixed crystals with smooth change in the lattice constant [6]. Similar constructions with graded composition layers were suggested and tested [7] for a deflection of particles. There is no need for the external force in such devices and they look very promising if considerable large and defects free quality crystal could be attained by molecular beam epitaxy (MBE) technique. A very simple method to bend the channels was recently proposed by a Russian-Italian collaboration[8]. In this method, small scratch on the side surface of the crystal causes a local deformation of the crystallographic planes down to a few hundred microns. The gradient of deformation creates the different curvature of adjacent channels. This can explain a peculiar focusing effect observed in series of experiment at Serpukhov and described in the reference[9], pp.118-122.

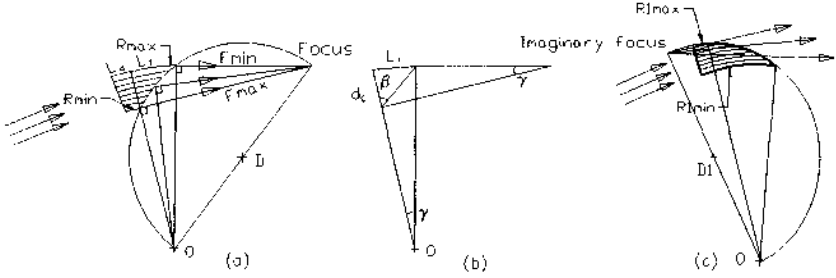

Fig. 1. Focusing (a) and defocusing (c) crystals for high energy particles. All angles and curvatures are exaggerated. (a)The front part, $L_d$, of the crystal is not needed for focusing and serves only as a deflection unit. (b)The angle of cutting, $\beta$, is calculated from focusing pencil 7 (Eq. 4).

A different approach from mentioned above is the use a regular bent crystal with constant curvature and parallel channels, but the end face of the crystal must be shaped in such a way that the tangent lines to the channels on that face are converged to a single point, see Fig. 1(a). In this case the center points of the channels on the end face of the crystal constitute a cylindrical surface. The diameter, $D$, of this surface is greater than curvature radii, $R_{min} < R < R_{max}$, of all channels. Up until now only this focusing geometry was successfully tested[10,11], and the focusing spot of $43\mu m$ was attained at the distance of 0.5m.

This paper examines the physical limits on focusing with a similar geometry and proposes a way to improve the focusing parameters. We restrict our consideration by one crystal with given aperture uniformly filled by a particle beam. The thorough statistical analysis of focusing in 1-D and 2-D geometries and multiple crystals will be considered elsewhere.

## 2   Focusing parameters and their limits

When an partical beam is exactly aligned with a set of planes, the particles are captured into channeling trajectories and directed along the bent atomic channels (Fig. 1(a)). At the exit of the crystal, each channel produces a particle microbeam (or more precisely subnanobeam because its size is $\sim 0.2nm$ for

(110) Si) which is directed along the tangent line. Each microbeam is incoherent to adjacent microbeams due to well known[12] equilibrium tendency inside the channels. The microbeam will travel a distance F, $F_{min} < F < F_{max}$, to the focal point. The radius-vector $R$ from the center of curvature $O$ to the output point of each channel is perpendicular to the center beam line and satisfy the equation

$$D^2 = R^2 + F^2, \tag{1}$$

where diameter of the cylindrical surface, D, is a constant for all channels, but the radius $R$ and focal length $F$ are slightly different. The first and last microbeams in the crystal cross section are described by the equations

$$D^2 = R_{min}^2 + F_{max}^2, \tag{2}$$
$$D^2 = R_{max}^2 + F_{min}^2, \tag{3}$$

and the relation between the angle of focusing pencil 7 and the cut angle $\beta$ is

$$\tan \beta = \gamma * \frac{R_{max}}{d_c}, \tag{4}$$

where $d_c = R_{max} - R_{min}$ is the thickness of the crystal. Three Si crystals with size 70 * 15 * 2mm were used in the experiment[10,11], and table 1 shows the angles $\beta, 7$ and focal length $F$ for the experimental set up.

Table 1
Parameters of bent crystals calculated from experiment [10,11]

| NO of crystal(*) | F(m) | $\gamma(\mu rad)$ | $L_f(mm)$ | $\tan(\beta)$ | $\beta(degree)$ |
|---|---|---|---|---|---|
| 1 | 3.5 | 571 | 1.54 | 0.7714 | 37.6 |
| 2 | 1.4 | 1428 | 3.86 | 1.929 | 62.6 |
| 3 | 0.5 | 4000 | 10.8 | 5.4 | 79.5 |

*The crystals are numerated as in ref.[11].

If $7 = L_f/R_{max} \ll 1$ and $F_{max}/R_{max} \gtrsim 7$ then the cylindrical surface in Fig. 1 could be made by cutting the whole crystal at the angle $\beta = \arctan(L_f/d_c)$ and bending the cutting surface over the cylinder with diameter D. Table 1 shows also the effective crystal length, $L_f$, participated in the focusing process.

The microbeam from each bent channel has a divergence $\Psi = (1 - R_c/R)\Psi_p$ (see, e.g. [13]), where $fy_p$ is the Lindhard critical angle for planar channeling, $R_c$ is the Tsyganov radius ($\Psi_p = 25\mu rad$, $R_c = 0.26m$ in ref.[10,11]). Obviously, the focal spot is proportional to the focal distance $F_{max}$. Fig. 1(a) shows that a crystal with parameters that satisfies Eq. (1) can only slide around the cylindrical surface. The smaller the distance $F_{max}$, the smaller the focal spot.

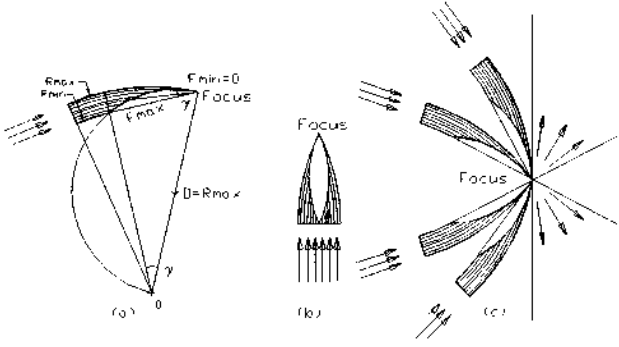

Fig. 2. (a)Focusing crystal with the smallest focus spot. (b) Intensity at the focus for one beam can be increased twofold using two adjacent symmetrical crystals. (c) Intensity at the focus for four beams.

The case with the minimum possible focal distance is shown in Fig 2(a)), where the focal spot lies almost on the tip of the crystal and $F_{min} = 0$. This geometry may give the smallest focal spot with maximum intensity, and it is interesting to consider it in more detail.

Eq. (3) with condition $F_{min} = 0$ leads to $D = R_{max}$, and the microbeam with maximum focal length must obey the equation

$$F_{max}^2 = R_{max}^2 - R_{min}^2. \tag{5}$$

Since the thickness of the crystal $d_c = R_{max} - R_{min}$, the maximum focal length in this case is given by

$$F_{max} = \sqrt{2d_c R_{max} - d_c^2}, \tag{6}$$

or in most practical cases

$$F_{max} \cong \sqrt{2d_c R}, \qquad R \sim R_{max} \sim R_{min}. \tag{7}$$

We may introduce the average focal distance $\overline{F} = (F_{max} + F_{min})/2$. For the crystal in Fig. 2(a), $\overline{F} = F_{max}/2$ and is proportional to the square root of the thickness $d_c$:

$$\overline{F} \cong \sqrt{\frac{d_c R}{2}}. \tag{8}$$

For the crystal bend radius $R = 2.7m$[10,11], the formula (8) gives $\overline{F} = 52$mm. This is a theoretical limit of possible focal lengths for given uniform bend and thickness of the crystal. The estimations of the full focal size and peak intensity can also easily be derived using Eq. (8):

$$\Delta_f \sim 2\Psi\overline{F} \equiv (1 - R_c/R)\Psi_p\sqrt{2d_cR}, \tag{9}$$

$$I_f \sim I_0 P_c^0 (1 - R_c/R)\frac{d_c}{\Delta_f} \equiv I_0 \frac{P_c^0}{\Psi_p}\sqrt{\frac{d_c}{2R}}, \tag{10}$$

where $I_0$ is the initial intensity, $P_c^0$ is the probability of particle capture in channeling mode by straight crystal, and $(1 - R_c/R)$ is the bent crystal factor [13]. It is important to note that the focal spot of the ideal crystalline lens and the focal peak intensity are proportional to the square root of the crystal thickness. The peak intensity is inversely proportional to the square root of the curvature radius. This does presume the crystal aperture is uniformly filled by the beam. It should be noted that Eqs. (9,10) are only estimations, accurate calculations will give the factor $f(R) \sim (1 - R_c/R)$ in Eq. (10), thus the intensity reaches a maximum at optimal radius $R = 3R_c$. If a crystal is cut according to the geometry drawn in Fig. 2(a) and with the experimental conditions of ref.[10,11], using Eq. (9, 10) one can estimate the best focal spot $\Delta_f \sim 3.0\mu m$, and magnification $I_f/(I_0 P_c^0) \sim 600$. It is 10 times better than was obtained in experiment [10,11]. The cutting angle $\beta$ in geometry of Fig. 2(a) must be $88.9 degree$. However, there are some obstacles to get extreme focal parameters because several defocusing factors may come into play. At small angles $(\pi/2 - \beta)$ the smoothness of the cylindrical surface is of great importance, the amorphous layer on that surface must be as thin as possible, and all adjacent channels must constitute the staircase with step widths equals to the interplaner space. Any deviation caused by surface roughness or mosaic blocks with widths about the length of channeling oscillation will produce a broadening of the focal spot and decrease the peak intensity. This could cause an abberation seen in [10,11] at focal length of 0.5m. There are techniques such as surface etching and argon polishing that can reduce the surface effects. Other sources of abberation to be considered for focusing are negative Gaussian curvature of bent crystals (anticlastic effect) and asymmetry of centripetal dechanneling in bent channels [9,13]. If the first can be avoided by special shape of the crystal, the last one is inevitable. There is also a possibility of asymmetry in coherent transitional scattering when the particles come out with small angles $(\pi/2 - \beta)$. For long crystal dechanneling is essential. In this case the thickness of the crystal $d_c$ should be substituted by $L_D/tan(\beta)$, where $L_D$ is the the dechanneling length. As a result the modified focal spot and intensity for long crystals are given by:

$$\Delta_f \sim (1 - R_c/R)\Psi_p\sqrt{\frac{2L_D R}{tan(\beta)}} \tag{11}$$

$$I_f \sim I_0 \frac{P_c^0}{\Psi_p}\sqrt{\frac{L_D}{2R\, tan(\beta)}}, \tag{12}$$

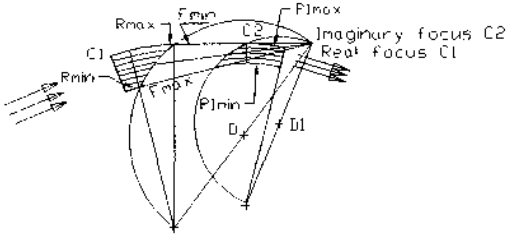

Fig. 3. Two crystal system for beam compressing. Front crystal has the focal distance greater than the second. As a result, a parallel beam is converted to a parallel beam with smaller cross profile.

On the contrary, there are many possibilities to increase the intensity at the focal spot. For example, the use of a combination of two symmetrical crystals with $L_d = 0$ filled by the one beam as it shown in Fig. 2(b) or the use of multiple beams, such as four beams as shown in Fig. 2(c). Such configuration can also be used for p-p colliding. At this point we should mention that the colliding applications, where the particle cycles through the crystal many times, suffer from serious problems of multiple scattering and radiation effects. Perhaps, the most practical applications of focusing crystals are in the area where the particle goes through the crystal only ones, e.g. two sequential crystals with concave surfaces for high efficient deflection of divergent beams [14,9]. If the front face of the crystal has a convex cylindrical form, the focal distances of such crystals might be considered negative (imaginary focus), see Fig. 1(c). Such crystals act in opposite way to the crystal in Fig. 1(a) and can convert a convergent beam pencil to a parallel beam. In general, the channeling trajectories are reversible and the reciprocity principle is held if the energy losses give negligible distortions [12]. So, reversing the beam (or changing input face to output) in short crystals always causes the opposite effect. Some interesting applications might come from the use of two crystals with real and imaginary foci. If their foci are coincided (telescopic configuration) the parallel beam is converted to another parallel beam with different width depending on the ratio of their focal distances. Fig 3 shows an example of compressing of the parallel beam, which can be helpful for accelerator and microbeam applications.

## 3 Conclusion

Summarizing, bent crystals with special input or output surface can focus and diverge beams of particles. The experimental results up to now are far from the physical limits of focusing. It is possible to improve the focal parameters one order of magnitude. However, special treatments must be done to keep the quality of the cylindrical output surfaces, which create the focal convergence. The system of two crystals offers several other interesting applications. One of them is the compression of cross section of the ion beam. Although it may suffer from a serious problem of losses the nowadays microbeam applications requires such devices.

## References

[1] E. N. Tsyganov, Some aspects of the mechanism of a charge particle penetration through a monocrystal, Fermilab,TM-682 (1976) 5.

[2] Y. N. Adishchev, et al., JETP lett. 30 (1979) 402.

[3] R. A. Carrigan, On the possible applications of the steering of charged particles by bent single crystal, Fermilab FN-80/45 (1980) 46.

[4] R. A. Carrigan, The application of channeling in bent crystals to charged particle beams, in: R. A. Carrigan, J. A. Ellison (Eds.), Relativistic Channeling. Plenum, New York, 1987, pp. 339-368.

[5] C. R. Sun, Aplication of channeling to particle physics, in: R. A. Carrigan, J. A. Ellison (Eds.), Relativistic Channeling. Plenum, New York, 1987, pp. 379–397.

[6] A. Schafer, W. Greiner, Possible use of mixed crystals to focus ion beam, J. Phys. G: Nulc. Part. Phys. 17 (1991) L217-L221.

[7] M. B. H. Brees, Beam bending using graded composition strained layers, Nucl. Inst. and Meth. B 132 (1997) 540–547.

[8] S. Bellucci, S. Bini, V. Biryukov, Y. Chesnokov, S. Dabagov, G. Giannini, V. Guidi, Y. Ivanov, V. Kotov, V. Maisheev, C. Malagu, G. Martinelli, A. Petrunin, V. Skorobogatov, M. Stefancich, D. Vincenzi, Experimental study for the feasibility of a crystalline undulator, arxiv.org/list/physics/0208028 (2002) 8.

[9] V. M. Biryukov, Y. A. Chesnokov, V. I. Kotov, Crystal Channeling and Its Application at High-Energy Accelerators, Springer, Verlag Berlin Heidelberg, 1997.

[10] A. S. Denisov, O. L. Fedin, M. A. Gordeeva, et al., First results from study of a 70 gev proton beam being focused by bent crystal, Nucl. Instr. and Meth. B 69 (1992) 382-384.

[11] V. I. Baranov, et al., Results of studying 70$gev$ proton beam focusing by bent crystal, in: XVth International Conference on High Energy Accelerators, Hamburg, Germany, July 20-24, 1992, pp. 128-130.

[12] J. Lindhard, Influence of crystal lattice on motion of energetic charged particles, Danske Vid. Selsk. Mat.-Fys. Medd. 34 (1965) no.14.

[13] A. M. Taratin, Particle channeling in a bent crystal, Phys. Particles Nuclei 29 (5) (1998) 437-462.

[14] V. I. Baranov, V. M. Biryukov, A. P. Bugorsky, Y. A. Chesnokov, V. I. Kotov, et al., Highly efficient deflection of a devergent beam by a bent single crystal, Nucl. Instr. and Meth. B 95 (1995) 449–451.